\documentclass[11pt,preprint]{aastex}

\newcommand{\etal}{{\it et~al.}}

\begin{document}

\title{Asteroid Diameters and Albedos from NEOWISE Reactivation Mission Years Four and Five}

\author{Joseph R. Masiero\altaffilmark{1}, A.K. Mainzer\altaffilmark{2}, J.M. Bauer\altaffilmark{3}, R.M. Cutri\altaffilmark{4}, T. Grav\altaffilmark{2,5}, E. Kramer\altaffilmark{1}, J. Pittichov\'{a}\altaffilmark{1}, S. Sonnett\altaffilmark{5}, E.L. Wright\altaffilmark{6}}

\altaffiltext{1}{Jet Propulsion Laboratory/California Institute of Technology, 4800 Oak Grove Dr., MS 183-301, Pasadena, CA 91109, USA, {\it Joseph.Masiero@jpl.nasa.gov}}
\altaffiltext{2}{Lunar and Planetary Laboratory, University of Arizona, Tucson, AZ, 85721, USA}

\altaffiltext{3}{University of Maryland, College Park, MD, 20742, USA} 
\altaffiltext{4}{California Institute of Technology, IPAC, 1200 California Blvd, Pasadena, CA 91125 USA}
\altaffiltext{5}{Planetary Science Institute, Tucson, AZ 85719 USA}
\altaffiltext{6}{University of California, Los Angeles, CA, 90095, USA} 

\begin{abstract}

The Near-Earth Object Wide-field Infrared Survey Explorer (NEOWISE)
spacecraft has been conducting a two-band thermal infrared survey to
detect and characterize asteroids and comets since its reactivation in
Dec 2013.  Using the observations collected during the fourth and
fifth years of the survey, our automated pipeline detected candidate
moving objects which were verified and reported to the Minor Planet
Center.  Using these detections, we perform thermal modeling of each
object from the near-Earth object and Main Belt asteroid populations
to constrain their sizes.  We present thermal model fits of asteroid
diameters for $189$ NEOs and $5831$ MBAs detected during the fourth
year of the survey, and $185$ NEOs and $5776$ MBAs from the fifth
year.  To date, the NEOWISE Reactivation survey has provided thermal
model characterization for $957$ unique NEOs.  Including all phases of
the original WISE survey brings the total to $1473$ unique NEOs that
have been characterized between 2010 and the present.

\end{abstract}

\section{Introduction}

The Near Earth Object Wide-field Infrared Survey Explorer
\citep[NEOWISE,][]{mainzer14} has been continuously surveying the sky
since 13 Dec 2013.  NEOWISE utilizes the Wide-field Infrared Survey
Explorer \citep[WISE,][]{wright10} satellite that was reactivated to
discover and characterize near-Earth asteroids in an effort to
quantify the risk they pose to Earth.  All NEOWISE images and extracted
source data from the first five years of the Reactivation survey are publicly
accessible via the NASA/IPAC Infrared Science Archive
(IRSA\footnote{https://irsa.ipac.caltech.edu}).  The content and
characteristics of NEOWISE data are described in \citet{cutri15}.

Observations of NEOs offer us the opportunity to study the smallest
asteroids as they pass close to the Earth, when they are significantly
easier to see.  These objects, having escaped from the Main Belt or
the Jupiter-family comet populations
\citep[e.g.][]{bottke02,granvik18}, let us probe the physics of the
formation and evolution of sub-kilometer-sized bodies.  NEOs also
represent a potential hazard to Earth, and thus survey and
characterization of them enables us to better quantify the chances of
impact and the dangers these objects pose.

The NEOWISE team has previously published the thermal modeling results
from the first three years of the Reactivation survey in
\citet{nugent15,nugent16,masiero17}. These fits, along with those from
previous survey phases, have been archived in the NASA Planetary Data
System \citep{mainzer19}.  Here, we perform the same analysis on the
data collected during the survey's fourth and fifth years (13 Dec 2016
to 12 Dec 2017, and 13 Dec 2017 to 12 Dec 2018, respectively).  At the
end of the NEOWISE reactivated survey, all asteroid thermal modeling
results will be delivered to the NASA Planetary Data System, to
augment the current archive from the original phases of the WISE
mission and the first three years of the NEOWISE Reactivation survey
currently archived there. The current mission plan is to operate
through June 2020, however the slower-than-expected evolution of the
spacecraft's orbit may allow for further useful survey lifetime, which
is currently being evaluated.

\section{Observations}

NEOWISE scans the sky along lines of constant ecliptic longitude,
recording images every 11 seconds, as the spacecraft orbits the Earth
in a 94-minute polar orbit.  The spacecraft was originally launched
onto a terminator-following orbit. Since then, as expected, the orbit
has gradually precessed off of the terminator to an average offset of
$\sim18-22^\circ$ during the survey's fourth and fifth
years\footnote{An illustration of this precession is shown in
    \citet{cutri15}, Sec I.2.b, Figure 8}.  On the evening side of
the orbit the spacecraft continues to survey at the zenith point with
respect to Earth, and thus at larger Solar elongations, but on the
morning side the telescope cannot point closer to the Sun and
therefore must maintain a pointing at Solar elongation of
$\sim90^\circ$, away from the local zenith point.  This
off-zenith pointing results in an increase in the heat load
on the telescope from the Earth that gradually raises the telescope
temperature over time.  As in the past, NEOWISE continues to toggle
its scan circles to avoid the Moon, speeding and slowing the
progression of the survey to avoid directly scanning over it.  NEOWISE
collects $\sim12$ detections per moving object over a span of $\sim30$
hours for objects near the ecliptic.  Objects closer to the ecliptic
poles can follow the survey region for long periods of time resulting
in longer sets of observations, while objects near the detection limit
may be detected fewer times as noise and light curve variations shift
them below the cutoff level.

Over the course of six months as the Earth orbits the Sun, NEOWISE
obtains images of the entire inertial sky.  As NEOs often have similar
orbits to the Earth, and thus long synodic periods, objects previously
undetected by NEOWISE regularly pass through the survey's field of
regard.  NEOWISE employs the WISE Moving Object Processing System
\citep[WMOPS,][]{mainzer11} to perform regular searches of the survey
data for new and known moving Solar System objects.  This is done
initially without incorporating any knowledge about the previous
discovery status of an object, enabling us to use the recovery of
previously known objects as a test of the efficiency of discovering
new ones \citep[see][]{mainzer11neo}.  WMOPS is run three times per
week, and all tracklets verified by our quality assurance process are
submitted to the Minor Planet Center
(MPC)\footnote{https://www.minorplanetcenter.net} for publication and
archiving.

WMOPS requires a minimum of 5 detections at a signal-to-noise ratio of
SNR$>4.5$, and has been shown to have an efficiency of $85\%-90\%$ for
bright objects within our selection requirements on number of
detections and motion vectors\citep{mainzer11neo}.  However, there are
some objects that are observed but will be missed due to low SNR, or
because they were moving too quickly through the field of regard and
were not seen a sufficient number of times, or because they had highly
curved motions on the plane of the sky that violate our
linearity-of-motion requirements for tracklet linking (a velocity
difference of 0.01 deg/day or a velocity angle change of 1 degree
between pairs of detections, which typically span 3 hours).  Searches
for known near-Earth objects found by other surveys with detections
not identified by WMOPS were carried out for the cryogenic WISE
mission data \citep{mainzer14tinyneo} and the first three years of the
NEOWISE reactived survey data \citep{masiero18tinyneo}, recovering
detections of $105$ and $116$ NEOs not found by WMOPS, respectively.
A similar search of the NEOWISE Reactivation Years 4 and 5 data is
underway and will be presented in future work.

The NEOWISE telescope uses a beamsplitter to collect images
simultaneously in the $3.4~\mu$m (W1) and $4.6~\mu$m (W2) bandpasses.
For objects with heliocentric distances near 1 AU, W2 is
generally dominated by thermal emission while W1 can be thermally
dominated or a mixture of thermal emission and reflected light
depending on the temperature of the object and how reflective the
object is at $3.4~\mu$m.  For more distant objects, e.g. Main Belt
asteroids (MBAs), W1 is almost always dominated by reflected light and
W2 can range from thermally-dominated to reflected-light-dominated
depending on the object's distance from the Sun and $4.6~\mu$m
reflectivity.  As a result, for the majority of detected NEOs we have
sufficient information to perform basic thermal modeling using
simplifying assumptions to reduce the number of variable parameters
(such as assuming the value for the beaming parameter and
  ratio of the infrared albedo to the visible albedo).  For MBAs,
conversely, only about half of the objects detected had sufficient
thermal emission to allow thermal modeling to set a constraint on the
diameter.  The remaining objects, which had significant contributions
of reflected light to both NEOWISE bandpasses, are not included in the
subsequent analysis.  Astrometric detections of them are still
recorded in the Minor Planet Center's database.

For more details on the survey and telescope, refer to the NEOWISE
Explanatory Supplement \citep{cutri15}, which is updated for each
annual NEOWISE data release.

\section{Thermal Modeling Technique}

The measured thermal flux from an asteroid depends on the object's
temperature, observing geometry, and size.  When enough astrometric
measurements are available to allow for the orbit to be constrained,
the distances to the Sun, Earth, and spacecraft as well as the
  phase angle at the time of observation will be sufficiently
well-known to contribute negligible error to the final thermal model
fit.  Thus, by employing a model of the thermal properties of the
surface, along with the known observational geometries, the diameter
of the asteroid can be constrained based on the measured flux in the
thermally dominated bands.  Using optical measurements from the
literature (in particular, the absolute $H$ magnitude published along
with the orbital information) the albedo of the asteroid can also be
constrained, however the uncertainty on this value depends on the
uncertainties on the diameter and the $H$ magnitude
\citep[cf.][]{masiero18}.

\subsection{Data}

The process for data extraction follows the same method used in
\citet{masiero17}.  To extract the data for use in thermal fitting, we
refer to the Minor Planet Center's Observations
Catalog\footnote{\it{http://minorplanetcenter.net/iau/ECS/MPCAT-OBS/MPCAT-OBS.html}},
which contains all observations of asteroids and comets submitted by
NEOWISE (observatory code C51) that were vetted and published by the
MPC.  We extracted all observations from C51 within survey Year 4 and
Year 5.  By using the MPC-accepted observations, we have a data set
that has initial source rejection done by the WMOPS pipeline as
well as subsequent checks on positional offsets by MPC that can flag
the occasional observation that was contaminated by cosmic rays or
other artifacts.  To obtain the fluxes associated with each detection
reported to the MPC we use the position-time measurements as an input
for the search of the NEOWISE Single-Exposure source database hosted
by IRSA, conducting a search for extracted sources within 5 arcsec of
the position and 5 secs of the MJD reported to the MPC.

NEOWISE source detection and photometry is carried out using the
expected PSF at that location on each detector simultaneously
\citep{cutri15}.  The quality of the fit between the PSF and the
identified source is recorded as a reduced $\chi^2$ value for each
bandpass.  We performed a filtering on the detections prior to using
them for thermal modeling based on their reduced $\chi^2$ of the fit
of the model PSF to the W2 detection (parameter $w2rchi2$), removing
any detection with $w2rchi2>5$.  This cut removes detections that may
be contaminated by cosmic rays or other spurious noise that could
potentially bias the fitted diameter.  We also remove from
consideration any object with an orbital arc shorter than 0.01 years,
as these objects received little-to-no ground based followup and thus
have uncertain orbits which will result in potentially incorrect
distance calculations and thermal modeling results.  This last cut
removed 3 NEOs and 15 MBAs from the Year 4 observation list, and 9
MBAs from the Year 5 list (no NEOs were removed by this cut in Year
5). Future recovery of these objects by other telescopes would enable
orbit fitting and thermal modeling, however at the moment the current
orbital knowledge is insufficient.

To constrain the optical albedos for objects as part of our thermal
modeling, we use the photometric $H_V$ absolute magnitude and $G$
slope parameter provided by the Minor Planet Center.  When available,
we updated the H-G parameters using the values published by
\citet{veres15} from the Pan-STARRS survey for objects that had phase
coverage $>1^\circ$ in those data \citep[cf.][for discussion on the
  effects of using these values]{masiero17}.  As these values are
based on a single, well-calibrated photometric system, they offer some
improvement over the values used by the MPC which incorporate a number
of different surveys with different levels of photometric calibration.
When not otherwise provided, we assume an uncertainty on $H$ of
$0.05~$mag for the Main Belt and $0.2~$ mag for the NEOs, and an
uncertainty on $G$ of $0.1$ based on our previous thermal modeling
experience.  As our sample of detected MBAs is primarily low-numbered
objects with long orbital arcs, the small assumed uncertainty on $H$
is appropriate.  Assuming a larger uncertainty on $H$ for MBAs can
allow a reflected light measurement in W1 to dominate the
least-squares fitting of that component of the model, and result in
poor matches to the published $H$ magnitude in some cases.

\subsection{NEATM}

For our fitting, we employ the Near-Earth Asteroid Thermal Model
\citep[NEATM,][]{harris98}.  This model provides a simple description
of the behavior of temperature across the surface of a spherical
asteroid, making use of a ``beaming parameter'' $\eta$ to consolidate
uncertainties in the assumed values of the physical properties and
differences between the model and actual temperature distribution.
Extreme values in the beaming parameter can also provide indications
of potentially unusual composition \citep[cf.][]{harris14}.  In all
cases, we used an assumed value for the beaming parameter based on the
distribution of fitted beaming values from the cryogenic NEOWISE
mission \citep{mainzer11neo,masiero11}.  For NEOs we assume
$\eta=1.4\pm0.5$, while for MBAs we assume $\eta=0.95\pm0.2$.  This
$1~\sigma$ uncertainty is used when conducting our Monte Carlo
analysis to propagate to the final uncertainty on the fitted
parameters diameter and albedo and is assumed to be normally
distributed around the mean value.

We note that in previous analyses
\citep[e.g.][]{nugent15,nugent16,masiero17} we allowed beaming to be a
fitted parameter for some NEOs where our observations indicated they
were likely to be thermally dominated in both W1 and W2 bands.  For
this work, our thermal modeling code was updated to Python 3.
Comparison of the output between the two versions shows that in the
vast majority of cases ($98.6\%$) the software converges to identical
solutions within the expected precision of the numerical routines.  In
a few cases when beaming was allowed to vary, the different code
versions could settle to distinct solutions.  This is because subtle
differences in initial conditions result in slightly different
estimates for the fraction of flux in W1 that is due to thermal
emission, which would then change whether the code allowed beaming to
vary or not.  It is important to note that in these cases our Monte
Carlo error analysis correctly captured the uncertainty on the
diameter solutions.  The different diameters were well within the
large resulting uncertainties.  As these are edge cases that straddle
fixed/fitted decision point in the code, and we have no independent
method of determining if beaming should be fitted or fixed, for the
current analysis we hold beaming fixed in all cases.

Previous work \citep{mainzer12,masiero12,nugent15,nugent16,masiero17}
has shown that the characteristic $1 \sigma$ diameter uncertainty for
the population of objects observed with W1 and W2 and fit with NEATM
is $\sim20\%$.  More detailed thermophysical models, which constrain
physical surface properties, can be used to perform multi-epoch fits
which can potentially offer improved diameter constraints when a range
of viewing geometries is available \citep[e.g.][]{alilagoa14,hanus18}.
However, these models take many orders of magnitude longer to run, and
are not likely to return improved results compared to the NEATM unless
observations covering a wide range of phase and distances are
available.  Further, these models require knowledge of the
  spin period and pole direction, or sufficient data to fit those
  parameters, which are not present for the majority of asteroids that
  have limited infrared and/or visible coverage.  As such, we provide
NEATM fits to all detected objects with sufficient data to better
understand the larger population and identify objects that may be of
interest for more detailed modeling in the future.

Our fitting procedure follows the method used in previous work
\citep[e.g.][]{masiero17}.  In summary, each photometric
observation from NEOWISE acts as a measurement to be fit by
  the Python least-squares fitting routine in the {\it scipy}
  package\citep{scipy}.  Observations include the position,
time, magnitudes and uncertainties in W1 and W2, and spacecraft
positions.  We require that an object have at least 3 measurements
with magnitude uncertainty $<0.25~$mag (SNR$~\sim4$) in a WISE band
for it to be used in fitting.  We only use an additional band
for fitting if the number of detections is more than $40\%$ of the
number in the band with the largest number of detections.  This
requirement is designed to remove potential contamination from cosmic
rays and background objects that may have been missed by other
filters.  It also results in a requirement that there are at least 3
detections for a second band to be used in the minimum case of a
5-detection tracklet (the lower limit produced by WMOPS).  The
published $H$ and $G$ visible photometric parameters are also included
as a measurement to be fitted by the least-squares fitter.  The
asteroid's orbit is used to determine Sun-to-object and
object-to-spacecraft distances as well as phase angle at each
observation time, which is used by NEATM to determine the temperature
distribution across the surface.  Specifically, we use a faceted
sphere made up of 288 facets in bands spaced at 15 degrees in
  latitude and calculate the temperature on each facet as well as the
resulting emission that would be observed.

Reflected light at visible wavelengths is constrained by the $H$
magnitude measurement.  To constrain the reflected light in the
NEOWISE bandpasses, we assume a ratio of albedos between the infrared
and visible of $1.5\pm0.5$ for MBAs and $1.6\pm1.0$ for NEOs, based on
the best fit values found during the cryogenic WISE mission for
objects where these parameters were fitted
\citep{mainzer11neo,masiero11}.  (These are different because
  the NEO population is not a random sample of the MBAs, but
  over-represents asteroids from the $\nu_6$ region.) In cases where
the W1 band is dominated by reflected light and has a very high SNR,
the assumed infrared-to-visible albedo ratio can result in the
least-squares minimizer finding a best fit solution where the
predicted $H$ magnitude (based on the fitted diameter and optical
albedo) does not match the measured value exactly.  Fits with large
deviations between the model and measured $H$ magnitudes are checked
to ensure the solutions are physically plausible.  In addition, fits
with visible albedos below $p_V<0.01$ or above $p_V=0.9$ are also
checked.  Previous work \citep[see][]{masiero17} found that the
majority of fits producing nonphysical or otherwise suspect results
occurred when more than $10\%$ of the flux in W2 was contributed by
reflected light based on best-fit parameters.  Following that work, we
discard these fits as unreliable, and do not include them in our final
tables.

The statistical uncertainty on each fitted parameter is determined by
performing 25 Monte Carlo trials, using uncertainties on each
measurement and the estimated uncertainties for assumed parameters.
Each trial draws a new value from a normal distribution around the
measured or assumed parameter, and conducts a least-squares fit to
those parameters.  The standard deviation of the population of all
Monte Carlo model fitted parameters is then taken as the $1 \sigma$
uncertainty.  These quoted uncertainties will only represent the
statistical component of the model fit, and do not account for
systematic offsets of the NEATM model with respect to reality.

\section{Results}

We present our model fits for NEOs and MBAs observed during Year 4 in
Table~\ref{tab.NEOfits4} and Table~\ref{tab.MBAfits4} respectively.
Fits for NEOs and MBAs observed during Year 5 are given in
Tables~\ref{tab.NEOfits5} and \ref{tab.MBAfits5} respectively.  Year 4
contains $214$ fits of $189$ unique NEOs, and $6658$ fits of $5831$
unique MBAs.  Year 5 contains $215$ fits for $185$ unique NEOs, and
$6600$ fits of $5776$ unique MBAs.  Each table gives the object's name
(in MPC-packed format), the measured $H$ and $G$ values used in the
process of fitting, the number of observations used in W1 and W2, the
orbital phase angle at the midpoint of the observations, along with
the best-fit diameter, the visible albedo, and beaming
parameter, with their associated uncertainties.  As we held beaming
fixed for all fits in this work, the beaming flag in the tables are
all set to $0$, but the flag is retained for easy comparison to
previously published results.  For objects that were seen at multiple
epochs in a given year, we present each fit as a separate entry in the
tables.  For objects that have non-spherical shapes, different epochs
can help constrain the true spherical equivalent diameter instead of
the projection-dependent results from a single epoch.  Alternately,
different epochs can provide insight into the thermal behavior of the
surface.  Thus, different diameter constraints from different epochs
of observation could be due to changing physical parameters, or simply
be a result of statistical noise.

\begin{figure}[ht]
\begin{center}
\includegraphics[scale=0.6]{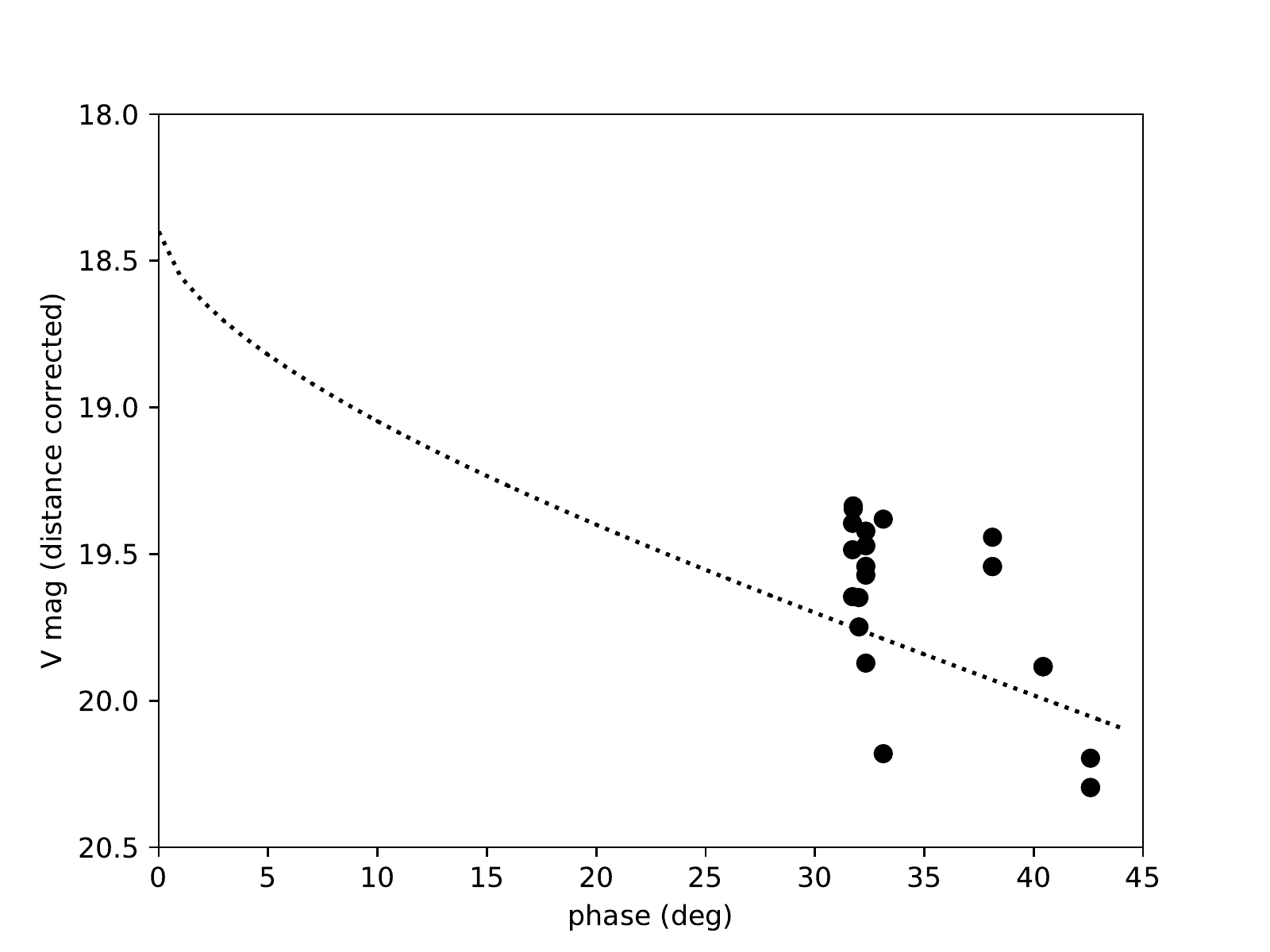}
\protect\caption{Distance-corrected magnitude measurements for
  asteroid 2018 KK$_2$ from the Minor Planet Center observation
  database are shown with black points.  The dashed line is the
  expected photometric behavior of an object with $H=18.4$ and
  $G=0.15$. Magnitudes were converted from the observed G and R bands
  assuming Solar colors of $V-R=0.36$ and $G-V=-0.14$
  \citep[][Gaia Data Release Documentation ({\it http://gea.esac.esa.int/archive/documentation/GDR2/})]{ramirez12}, compatible with a flat-spectral slope expected for low-albedo C-type objects.}
\label{fig.KK2}
\end{center}
\end{figure}

We note that one object in the Year 5 NEO table, 2018 KK$_2$, has a
best-fit albedo $p_V<0.01$ despite our attempts at filtering or
changing assumed parameters.  This Amor-class NEO has an orbital arc
spanning $\sim3$ months with observations that span $\sim10^\circ$ of
phase, however the scatter in the photometry means that the published
$H$ value is not necessarily well-constrained (see
Figure~\ref{fig.KK2}).  The NEOWISE observations of this object
occurred at a phase of $\alpha=30^\circ$, so a poorly constrained G
value won't have as large an effect on the predicted brightness at the
time of our observations.  An underestimated brightness from an $H$
value that was too large would drive the albedo to artificially low
values.  The unphysically low albedo is then likely the result of a
combination of poor $H$ fit and statistical uncertainty on the size
measurement, possibly combined with light curve variations.  We
include the best fit as reported by our model in the results table.
While the diameter should be reliable to the quoted errors, caution
should be used regarding the interpretation of the albedo for this
object.  This highlights the impact that uncertainty on the optical
measurements has on our ability to determine albedos, and shows the
need for improved $H$ and $G$ determinations for all objects from a
photometrically calibrated survey \citep[e.g.][]{juric02,veres15}.

\begin{figure}[ht]
\begin{center}
\includegraphics[scale=0.6]{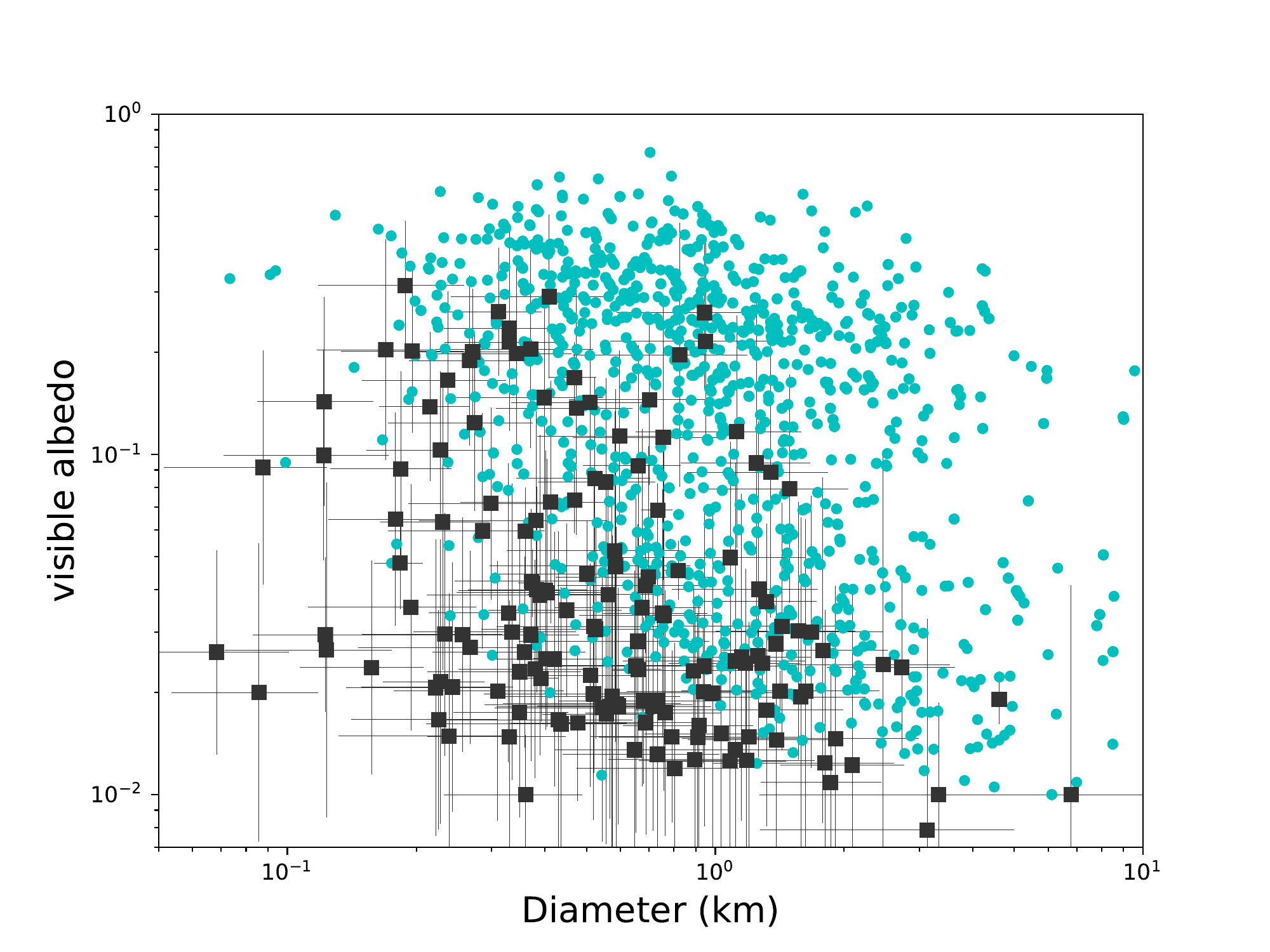}
\protect\caption{Comparison of fitted diameters and albedos for all
  near-Earth objects observed (cyan circles) and discovered (black
  squares) by NEOWISE during the first five years of the reactivation
  survey (Dec 2013 to Dec 2018) by the WMOPS pipeline.  The majority
  of objects discovered by NEOWISE tend to have albedos below $10\%$
  and diameters larger than a few hundred meters, filling in a region
  of phase space missed by other surveys \citep[cf.][figure
    14]{mainzer11neo}.  Error bars on previously known objects are
  omitted for clarity, but are of comparable size to the uncertainties
  on the NEOWISE discoveries.}
\label{fig.diamalb}
\end{center}
\end{figure}

We show a plot of diameter and albedo for all near-Earth objects
detected by NEOWISE from Dec 2013 to Dec 2018 in
Figure~\ref{fig.diamalb}.  Objects discovered by NEOWISE show a
preference for being low albedo, with many of them being larger than
$200~$m in diameter.  This population of objects is more likely to be
missed by the visible light ground-based surveys due to
albedo-dependent selection effects inherent in those systems.  Thus,
while NEOWISE is primarily a NEO-characterization mission, it fills an
important part of phase space in the current suite of near-Earth
object discovery surveys.

\begin{table}[ht]
\begin{center}
\caption{Thermal model fits for NEOs detected in the fourth year of the
  NEOWISE survey.  Table 1 is published in its entirety
  in the electronic edition; a portion is shown here for guidance
  regarding its form and content.}
\vspace{1ex}
\noindent
\begin{tabular}{cccccccccc}
\tableline
  Name  &  H$^\dagger$  &   G   &   Diameter  &  $p_V^{\dagger\dagger}$  &   beaming$^{\dagger\dagger\dagger}$  &  n$_{W1}$  &n$_{W2}$ & phase & Fitted \\
  & (mag) & & (km) &&&&& (deg) & Beaming? \\
  \tableline

  01864 & 14.85 &  0.15 &    2.73 $\pm$   0.79 & 0.271 (+0.205/-0.117) & 1.40 $\pm$ 0.50 &   5 &   5 & 54.84 & 0\\
  02102 & 16.00 &  0.15 &    1.53 $\pm$   0.56 & 0.298 (+0.257/-0.138) & 1.40 $\pm$ 0.50 &   8 &   8 & 58.12 & 0\\
  02329 & 14.50 &  0.15 &    4.17 $\pm$   1.72 & 0.148 (+0.147/-0.074) & 1.40 $\pm$ 0.50 &  12 &  12 & 38.45 & 0\\
  03122 & 14.10 &  0.15 &    4.21 $\pm$   1.12 & 0.351 (+0.234/-0.140) & 1.40 $\pm$ 0.50 &  19 &  19 & 79.28 & 0\\
  03122 & 14.10 &  0.15 &    4.28 $\pm$   1.27 & 0.346 (+0.237/-0.141) & 1.40 $\pm$ 0.50 &  33 &  33 & 73.15 & 0\\
  03352 & 15.80 &  0.15 &    1.55 $\pm$   0.43 & 0.274 (+0.203/-0.117) & 1.40 $\pm$ 0.50 &  17 &  18 & 50.01 & 0\\
  03752 & 15.30 &  0.15 &    2.48 $\pm$   0.92 & 0.238 (+0.208/-0.111) & 1.40 $\pm$ 0.50 &  14 &  16 & 48.65 & 0\\
  04179 & 15.30 &  0.10 &    2.64 $\pm$   1.05 & 0.254 (+0.240/-0.123) & 1.40 $\pm$ 0.50 &  10 &  10 & 61.09 & 0\\
  04197 & 14.60 &  0.15 &    3.67 $\pm$   1.47 & 0.155 (+0.149/-0.076) & 1.40 $\pm$ 0.50 &   5 &   5 & 58.95 & 0\\
  05653 & 16.20 &  0.15 &    1.60 $\pm$   0.63 & 0.261 (+0.246/-0.127) & 1.40 $\pm$ 0.50 &  21 &  24 & 50.25 & 0\\
  
\hline
\end{tabular}
\label{tab.NEOfits4}
$^\dagger$Measured H used as input for the modeling; the model-output H value can be found using the output diameter, albedo, and the equation $D = 1329*10^{H/-5}/\sqrt{p_V}$\\
$^{\dagger\dagger}$Albedo uncertainties are symmetric in log-space as the error is dominated by the uncertainty on $H$; the asymmetric linear equivalents of the $1 \sigma$ log-space uncertainties are presented here.\\
$^{\dagger\dagger\dagger}$Assumed constant value; not fit.

\end{center}
\end{table}

\begin{table}[ht]
\begin{center}
\caption{Thermal model fits for MBAs detected in the fourth year of the
  NEOWISE survey.  Table 2 is published in its entirety
  in the electronic edition; a portion is shown here for guidance
  regarding its form and content.}
\vspace{1ex}
\noindent
\begin{tabular}{cccccccccc}
\tableline
  Name  &  H$^\dagger$  &   G   &   Diameter  &  $p_V^{\dagger\dagger}$  &   beaming$^{\dagger\dagger\dagger}$  &  n$_{W1}$  &n$_{W2}$ & phase & Fitted \\
  & (mag) & & (km) &&&&& (deg) & Beaming? \\
  \tableline

  00010 &  5.43 &  0.15 &  438.31 $\pm$ 144.12 & 0.046 (+0.072/-0.028) & 0.95 $\pm$ 0.20 &   7 &   8 & 20.83 & 0\\
  00013 &  6.74 &  0.15 &  197.47 $\pm$  59.09 & 0.057 (+0.039/-0.023) & 0.95 $\pm$ 0.20 &   7 &   7 & 23.90 & 0\\
  00019 &  7.13 &  0.10 &  227.74 $\pm$  68.19 & 0.034 (+0.024/-0.014) & 0.95 $\pm$ 0.20 &   4 &   5 & 25.65 & 0\\
  00025 &  7.83 &  0.15 &   97.99 $\pm$  18.06 & 0.196 (+0.080/-0.057) & 0.95 $\pm$ 0.20 &  10 &  10 & 29.82 & 0\\
  00031 &  6.74 &  0.15 &  302.09 $\pm$ 113.80 & 0.035 (+0.032/-0.017) & 0.95 $\pm$ 0.20 &  10 &  10 & 23.70 & 0\\
  00034 &  8.51 &  0.15 &  116.40 $\pm$  47.83 & 0.038 (+0.037/-0.019) & 0.95 $\pm$ 0.20 &  10 &  10 & 22.26 & 0\\
  00046 &  8.36 &  0.06 &  124.71 $\pm$  28.64 & 0.052 (+0.044/-0.024) & 0.95 $\pm$ 0.20 &   8 &   8 & 21.88 & 0\\
  00050 &  9.24 &  0.15 &   87.30 $\pm$  30.40 & 0.035 (+0.028/-0.016) & 0.95 $\pm$ 0.20 &   6 &   9 & 29.67 & 0\\
  00051 &  7.35 &  0.08 &  134.63 $\pm$  36.55 & 0.090 (+0.055/-0.034) & 0.95 $\pm$ 0.20 &  13 &  13 & 25.93 & 0\\
  00051 &  7.35 &  0.08 &  128.44 $\pm$  31.02 & 0.087 (+0.051/-0.032) & 0.95 $\pm$ 0.20 &  12 &  12 & 22.62 & 0\\
   
\hline
\end{tabular}
\label{tab.MBAfits4}
$^\dagger$Measured H used as input for the modeling; the model-output H value can be found using the output diameter, albedo, and the equation $D = 1329*10^{H/-5}/\sqrt{p_V}$\\
$^{\dagger\dagger}$Albedo uncertainties are symmetric in log-space as the error is dominated by the uncertainty on $H$; the asymmetric linear equivalents of the $1 \sigma$ log-space uncertainties are presented here.\\
$^{\dagger\dagger\dagger}$Assumed constant value; not fit. 

\end{center}
\end{table}

\begin{table}[ht]
\begin{center}
\caption{Thermal model fits for NEOs detected in the fifth year of the
  NEOWISE survey.  Table 3 is published in its entirety
  in the electronic edition; a portion is shown here for guidance
  regarding its form and content.}
\vspace{1ex}
\noindent
\begin{tabular}{cccccccccc}
\tableline
  Name  &  H$^\dagger$  &   G   &   Diameter  &  $p_V^{\dagger\dagger}$  &   beaming$^{\dagger\dagger\dagger}$  &  n$_{W1}$  &n$_{W2}$ & phase & Fitted \\
  & (mag) & & (km) &&&&& (deg) & Beaming? \\
  \tableline

  00719 & 15.50 &  0.15 &    2.59 $\pm$   0.81 & 0.189 (+0.175/-0.091) & 1.40 $\pm$ 0.50 &  16 &  16 & 46.48 & 0\\
  01627 & 13.20 &  0.60 &    8.99 $\pm$   3.31 & 0.129 (+0.112/-0.060) & 1.40 $\pm$ 0.50 &  23 &  23 & 41.56 & 0\\
  01627 & 13.20 &  0.60 &    9.01 $\pm$   2.36 & 0.127 (+0.075/-0.047) & 1.40 $\pm$ 0.50 &  25 &  27 & 52.18 & 0\\
  01627 & 13.20 &  0.60 &    5.96 $\pm$   1.98 & 0.167 (+0.130/-0.073) & 1.40 $\pm$ 0.50 &   7 &   7 & 54.86 & 0\\
  01916 & 14.93 &  0.15 &    2.77 $\pm$   1.07 & 0.212 (+0.217/-0.107) & 1.40 $\pm$ 0.50 &   8 &   8 & 43.30 & 0\\
  03552 & 12.90 &  0.15 &   26.89 $\pm$  12.58 & 0.028 (+0.032/-0.015) & 1.40 $\pm$ 0.50 &  12 &  13 & 35.31 & 0\\
  04596 & 16.30 &  0.15 &    2.10 $\pm$   0.74 & 0.173 (+0.144/-0.079) & 1.40 $\pm$ 0.50 &  11 &  11 & 45.89 & 0\\
  05797 & 18.70 &  0.15 &    0.45 $\pm$   0.16 & 0.288 (+0.274/-0.141) & 1.40 $\pm$ 0.50 &   0 &   7 & 48.69 & 0\\
  09856 & 17.40 &  0.15 &    0.91 $\pm$   0.39 & 0.245 (+0.258/-0.125) & 1.40 $\pm$ 0.50 &   8 &   8 & 66.67 & 0\\
  09856 & 17.40 &  0.15 &    0.94 $\pm$   0.35 & 0.252 (+0.221/-0.118) & 1.40 $\pm$ 0.50 &   8 &   8 & 53.74 & 0\\
    
\hline
\end{tabular}
\label{tab.NEOfits5}
$^\dagger$Measured H used as input for the modeling; the model-output H value can be found using the output diameter, albedo, and the equation $D = 1329*10^{H/-5}/\sqrt{p_V}$\\
$^{\dagger\dagger}$Albedo uncertainties are symmetric in log-space as the error is dominated by the uncertainty on $H$; the asymmetric linear equivalents of the $1 \sigma$ log-space uncertainties are presented here.\\
$^{\dagger\dagger\dagger}$Assumed constant value; not fit.

\end{center}
\end{table}

\begin{table}[ht]
\begin{center}
\caption{Thermal model fits for MBAs detected in the fifth year of the
  NEOWISE survey.  Table 4 is published in its entirety
  in the electronic edition; a portion is shown here for guidance
  regarding its form and content.}
\vspace{1ex}
\noindent
\begin{tabular}{cccccccccc}
\tableline
  Name  &  H$^\dagger$  &   G   &   Diameter  &  $p_V^{\dagger\dagger}$  &   beaming$^{\dagger\dagger\dagger}$  &  n$_{W1}$  &n$_{W2}$ & phase & Fitted \\
  & (mag) & & (km) &&&&& (deg) & Beaming? \\
  \tableline

  00013 &  6.74 &  0.15 &  219.07 $\pm$  78.46 & 0.053 (+0.045/-0.024) & 0.95 $\pm$ 0.20 &   9 &   9 & 20.42 & 0\\
  00013 &  6.74 &  0.15 &  235.06 $\pm$  73.14 & 0.047 (+0.037/-0.021) & 0.95 $\pm$ 0.20 &   5 &   5 & 22.57 & 0\\
  00019 &  7.13 &  0.10 &  190.24 $\pm$  57.83 & 0.057 (+0.042/-0.024) & 0.95 $\pm$ 0.20 &  12 &  12 & 21.41 & 0\\
  00034 &  8.51 &  0.15 &  105.98 $\pm$  30.95 & 0.041 (+0.027/-0.016) & 0.95 $\pm$ 0.20 &   9 &  10 & 23.85 & 0\\
  00034 &  8.51 &  0.15 &  105.52 $\pm$  30.53 & 0.042 (+0.044/-0.021) & 0.95 $\pm$ 0.20 &  19 &  19 & 21.63 & 0\\
  00038 &  8.32 &  0.15 &   87.36 $\pm$  36.65 & 0.061 (+0.062/-0.031) & 0.95 $\pm$ 0.20 &   9 &  10 & 23.89 & 0\\
  00041 &  7.12 &  0.10 &  179.21 $\pm$  56.70 & 0.053 (+0.043/-0.024) & 0.95 $\pm$ 0.20 &  17 &  17 & 22.72 & 0\\
  00045 &  7.46 &  0.07 &  138.44 $\pm$  40.24 & 0.076 (+0.085/-0.040) & 0.95 $\pm$ 0.20 &  11 &  11 & 21.47 & 0\\
  00045 &  7.46 &  0.07 &  175.21 $\pm$  44.60 & 0.058 (+0.058/-0.029) & 0.95 $\pm$ 0.20 &  13 &  13 & 21.70 & 0\\
  00046 &  8.36 &  0.06 &  118.09 $\pm$  29.19 & 0.050 (+0.028/-0.018) & 0.95 $\pm$ 0.20 &   3 &   4 & 29.06 & 0\\

\hline
\end{tabular}
\label{tab.MBAfits5}
$^\dagger$Measured H used as input for the modeling; the model-output H value can be found using the output diameter, albedo, and the equation $D = 1329*10^{H/-5}/\sqrt{p_V}$\\
$^{\dagger\dagger}$Albedo uncertainties are symmetric in log-space as the error is dominated by the uncertainty on $H$; the asymmetric linear equivalents of the $1 \sigma$ log-space uncertainties are presented here.\\
$^{\dagger\dagger\dagger}$Assumed constant value; not fit.

\end{center}
\end{table}

\clearpage

\section{Accuracy of the NEATM Thermal Modeling}

\begin{figure}[ht]
\begin{center}
\includegraphics[scale=0.5]{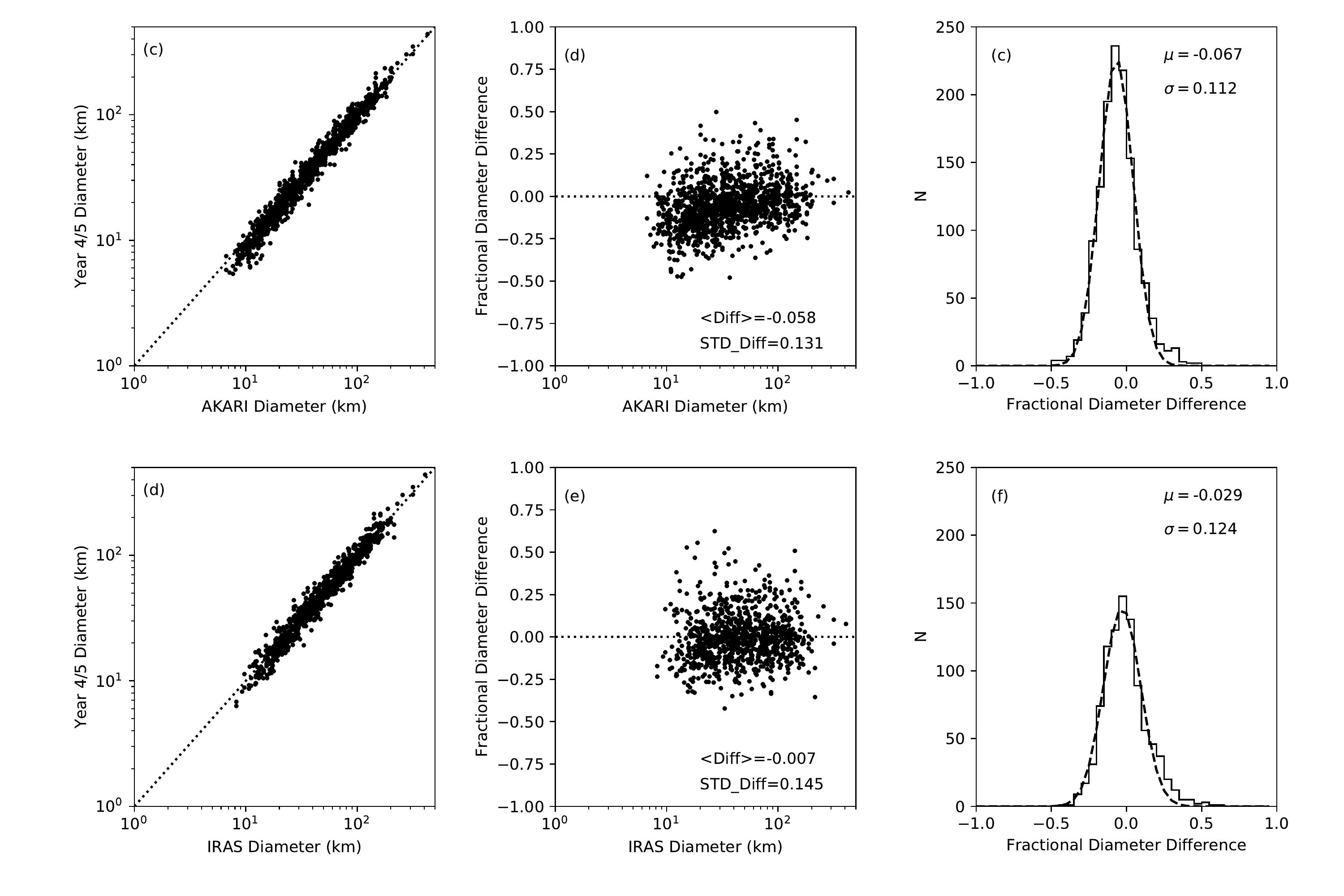} 
\protect\caption{Main Belt asteroid diameter fits from the NEOWISE
  Reactivation Years 4 and 5 data compared to diameters derived from
  AKARI measurements \citep[][panel (a)]{usui14} and IRAS
  \citep[][panel (d)]{irasPDS}.  Dotted lines show a 1:1 relationship.
  We show the fractional difference in fits against the comparison
  diameter ((year 4/5 - comparison)/comparison; panels (b) and (e))
  for each comparison set.  We also show the histogram of the
  fractional differences (panels (c) and (f)) along with the best-fit
  Gaussian to the fractional difference distribution and its mean
  ($\mu$) and standard deviation ($\sigma$).}
\label{fig.akariiras}
\end{center}
\end{figure}

As discussed in \citet{wright19}, the diameters derived by NEOWISE
have a characteristic $1 \sigma$ uncertainty in effective spherical
diameter of $\sim10\%$ for objects with sufficient data to fit
multiple thermal bands.  This was found through comparisons between
diameter fits from NEOWISE and diameters determined by the IRAS
satellite \citep{irasPDS}. \citet{usui14} found a similar result for
comparisons between the cryogenic NEOWISE fits and AKARI.  These works
focused on data from the cryogenic mission, so an independent check of
the fits based on 2-band Reactivation data is appropriate.  We show
the comparison between the NEOWISE Reactivation survey years 4 and 5
diameters and the diameters from the IRAS and AKARI data (for objects
with more than 5 detections to reduce selection effect biases) in
Figure~\ref{fig.akariiras}.

To additionally verify the diameters published here, we perform a
comparison of the Years 4 and 5 fits to previously published results.
For these comparisons, we use three primary datasets: a collection of
published radar reflection sizes
\citep{hudson94,magri99,shevchenko06,magri07,shepard10,naidu15} and
occultation timing chords \citep{durech11,occul_PDS}; diameters from
the fully cryogenic NEOWISE dataset which had fitted beaming
parameters; and diameters from the NEOWISE Reactivation Years 1-3.
NEOWISE diameters are drawn from the compilation in the NASA Planetary
Data System \citep[Version 2,][]{mainzer19}.

\begin{figure}[ht]
\begin{center}
\includegraphics[scale=0.5]{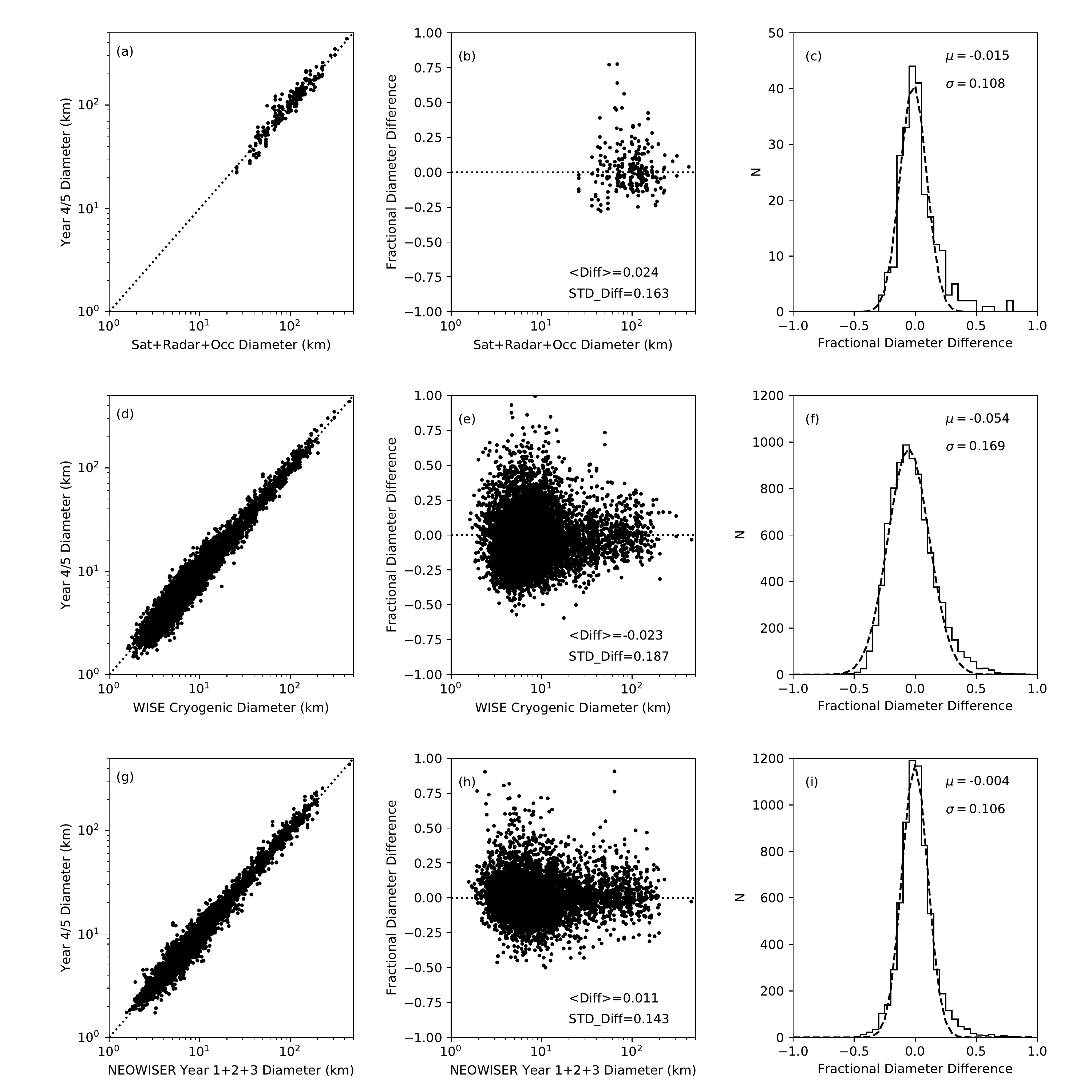} 

\protect\caption{Main Belt asteroid diameter fits from the NEOWISE
  Reactivation Years 4 and 5 data compared to diameters derived from radar and
  occultation measurements (panel (a)), NEOWISE fully cryogenic data
  (panel (d)), and NEOWISE Reactivation Years 1-3 data (panel(g)).  Dotted
  lines show a 1:1 relationship.  We show the fractional difference in
  fits against the comparison diameter ((year 4/5 -
  comparison)/comparison; panels (b), (e), (h)) for each comparison
  set.  We also show the histogram of the fractional differences
  (panels (c), (f), (i)) along with the best-fit Gaussian to the
  fractional difference distribution and its mean ($\mu$) and standard
  deviation ($\sigma$).}
\label{fig.mbacomp}
\end{center}
\end{figure}

\begin{figure}[ht]
\begin{center}
\includegraphics[scale=0.5]{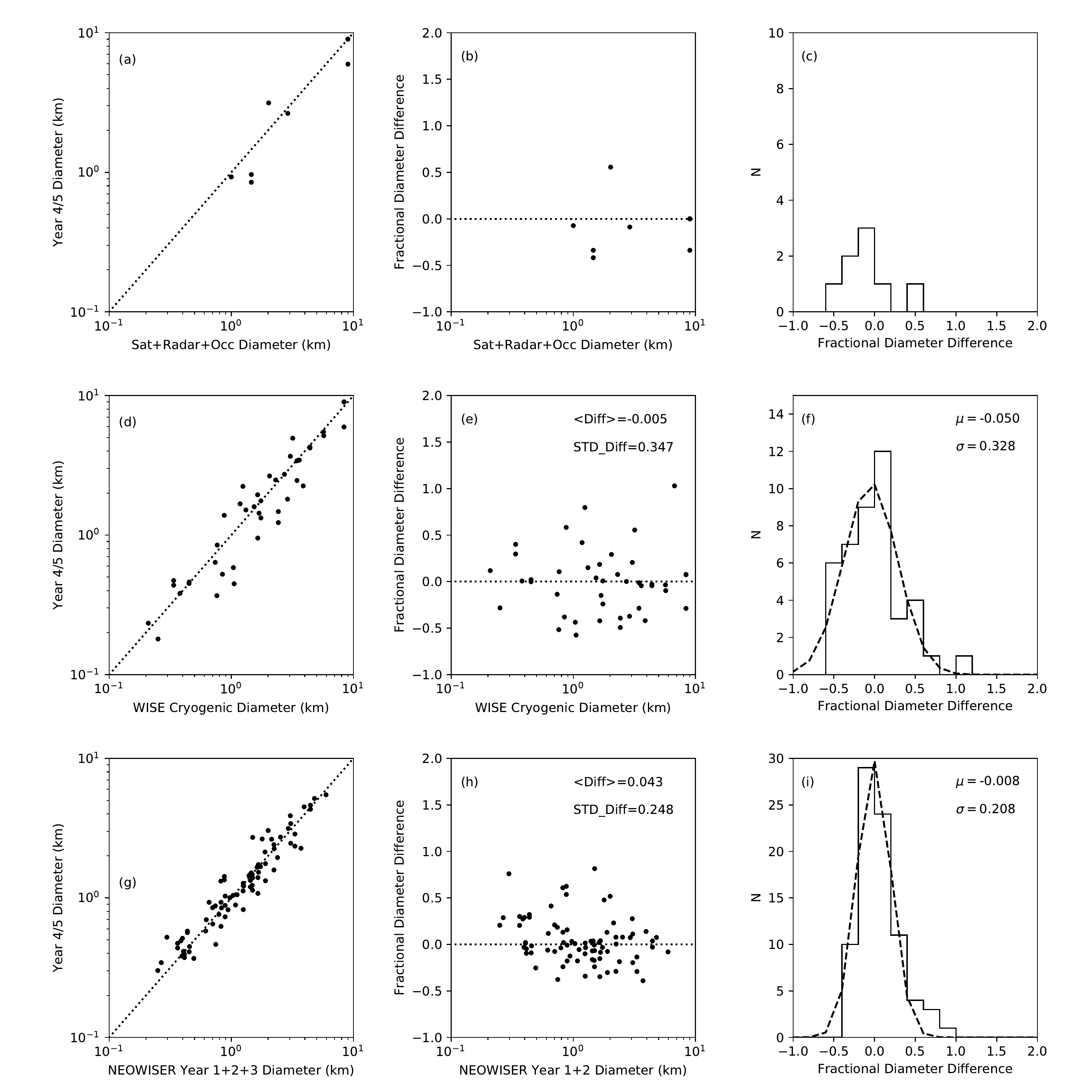} \protect\caption{The
  same as Figure~\ref{fig.mbacomp}, but for near-Earth objects
  observed during Years 4 and 5 that also were present in the
  comparison datasets.}
\label{fig.neocomp}
\end{center}
\end{figure}

Figure~\ref{fig.mbacomp} shows the comparison between the Main Belt
asteroid fits published here with previously published values, while
Figure~\ref{fig.neocomp} shows the near-Earth objects. For the Main
Belt population we find that the best-fit Gaussian to the diameter
deviations for the population shows $1 \sigma$ spreads of $11-17\%$
with systematic offsets of no more than a few percent.  The comparison
to the NEOWISE cryogenic diameters shows a $\sim5\%$ offset for the
population, with a comparable offset seen in the AKARI and IRAS
comparisons.  This offset is not seen in the comparison to the earlier
NEOWISE reactivation diameters, so indicates a shift between the
cryogenic and reactivation fits.  This offset, however, is within the
$10\%$ minimum systematic uncertainty we assume for our implementation
of our thermal model \citep{mainzer11cal}.

A few objects in our comparison of Main Belt sizes show large
deviations between the thermal modeled diameter and the size measured
by occultations \citep{occul_PDS}, with the thermal diameter being
much larger.  The largest outliers in our comparison are (90) Antiope
[81 vs 127 km], (415) Palatia [55 vs 98 km], and (431) Nephele [68 vs
  112 and 121 km at two different epochs].  All of these agree well
with diameters obtained through thermal modeling in other epochs of
NEOWISE \citep{mainzer19}.  For Antiope, the occultation diameter is
the size of the primary only of the equal-mass binary, so the
difference from size when assuming a single sphere (as we do in our
diameter calculation) is understood.  Palatia is a lower-confidence
(U=2) occultation, so not covered completely by chords.  For Nephele,
the chord coverage looks sufficient to constrain the full shape, so
this perhaps is another case of an equal-mass binary where only one
component was picked up by the occultations.

The near-Earth population shows a larger Gaussian spread of
$\sim20-30\%$, but uses fewer comparison objects because the NEO
population is smaller, and due to changing viewing geometry objects
are not as likely to be re-detected at different epochs.  This larger
spread for the NEOs may be an indication that the population is more
non-spherical than MBAs, or may simply be due to a combination of
statistical noise and the limitations of the NEATM model at high phase
angles, as discussed by \citet{mommert18}.  We do not fit a Gaussian
to the comparison between NEOs and non-infrared diameter sources due
to the small number of measurements in this dataset ($N=8$).

\clearpage

\section{Conclusions}

We present thermal model fits to near-Earth objects and Main Belt
asteroids detected during the fourth and fifth years of the
reactivated NEOWISE survey.  Included are $214$ fits of $189$ unique
NEOs and $6658$ fits of $5831$ MBAs from Year 4, and $215$ fits of
$185$ unique NEOs and $6600$ fits of $5776$ MBAs from Year 5.  We
follow the data quality restrictions used for the fits to the NEOWISE
Year 3 data \citep{masiero17}, in particular rejecting fits with
$>10\%$ modeled reflected light in the W2 band as unstable solutions.
This results in a large number of detected MBAs being rejected from
the thermal fit results, but improves the fit reliability.  This cut
will introduce a strong bias against high albedo objects in
the list of Main Belt objects characterized.

We find that the diameter fits for Main Belt asteroids have a
characteristic $1 \sigma$ uncertainty of $\sim15\%$ compared to other
data sets (assuming a Gaussian distribution), while NEOs show a larger
uncertainty of $\sim20-30\%$, consistent with what we have found in
previous years, though this is based on a smaller comparison set.
Thus there is no apparent degradation in the quality of the NEOWISE
data as the survey has continued.

NEOWISE has provided thermal model characterization of $957$ unique
NEOs during the Reactivation mission, bringing the total number of
NEOs characterized from all mission phases to $1473$, including NEOs
detected automatically by our WMOPS pipeline and those recovered later
using the IRSA moving object search tools.  The NEOWISE survey
continues into its sixth year of operation.  The spacecraft has
precessed from its original terminator-following orbit, though the
rate of precession has been slower than expected due to the low levels
of Solar activity in the last few years.  Eventually this precession
will force an end to the mission, though it is difficult to predict
the exact timing of when data quality will diminish.  NEOWISE data
continue to provide an important resource for discovering and
characterizing asteroids and comets.

\section*{Acknowledgments}

The authors would like to thank the two anonymous referees for their
helpful comments that improved this manuscript.  This publication
makes use of data products from the Wide-field Infrared Survey
Explorer, which is a joint project of the University of California,
Los Angeles, and the Jet Propulsion Laboratory/California Institute of
Technology, funded by the National Aeronautics and Space
Administration.  This publication also makes use of data products from
NEOWISE, which is a project of the Jet Propulsion
Laboratory/California Institute of Technology, funded by the Planetary
Science Division of the National Aeronautics and Space Administration.
The research was carried out at the Jet Propulsion Laboratory,
California Institute of Technology, under a contract with the National
Aeronautics and Space Administration (80NM0018D004).  This research
has made use of data and services provided by the International
Astronomical Union's Minor Planet Center.  This publication uses data
obtained from the NASA Planetary Data System (PDS).  This research has
made use of the NASA/IPAC Infrared Science Archive,which is funded by
the National Aeronautics and Space Administration and operated by the
California Institute of Technology.  This research has made extensive
use of the {\it numpy}, {\it scipy}, and {\it matplotlib} Python
packages. Based on observations obtained at the Gemini Observatory,
which is operated by the Association of Universities for Research in
Astronomy, Inc., under a cooperative agreement with the NSF on behalf
of the Gemini partnership: the National Science Foundation (United
States), the National Research Council (Canada), CONICYT (Chile),
Ministerio de Ciencia, Tecnolog\'{i}a e Innovaci\'{o}n Productiva
(Argentina), and Minist\'{e}rio da Ci\^{e}ncia, Tecnologia e
Inova\c{c}\~{a}o (Brazil).  The authors also acknowledge the efforts
of worldwide NEO followup observers who provide time-critical
astrometric measurements of newly discovered NEOs, enabling object
recovery and computation of orbital elements.  Many of these efforts
would not be possible without the financial support of the NASA
Near-Earth Object Observations Program, for which we are grateful.

\end{document}